\documentstyle[11pt,appbx3,epsfig]{article}
\newcommand{\slsh}[1]{\, {\not {\! #1}}}
\newcommand{\oabs}[1]{|\vec{#1} \makebox[0.1em]{} |}
\newcommand{\oabsq}[1]{{|\vec{#1} \makebox[0.1em]{} |}^2}
\begin{document}
\title{CONSISTENT EFFECTIVE FIELD--THEORETIC\\
       TREATMENT OF RESONANCES\\
       WITH NON ZERO WIDTH}
\author{Frieder Kleefeld
\address{
Inst.\ for Theoretical Physics III, University of Erlangen-N\"urnberg\\
Staudtstr. 7, 91058 Erlangen, Germany\\
e-mail: {\tt kleefeld@theorie3.physik.uni-erlangen.de}}
}
\setcounter{page}{981}
\maketitle

\begin{abstract}
A systematic field theoretic
formalism for treating Fermionic resonances with non zero width is given. The
implication of unitarity to coupling constants of non local interactions is
shown. The extension of the formalism to Bosonic resonance field operators is 
straight forward. \\[5mm]
\end{abstract}

\section{Field theoretic effective model for one resonance}

It is well known (see e.g.\ [1], p.\ 1 ff), that at energies close to a resonance the propagator of a theory
is dominated by its singular part, which has the following separable structure:
$G(E)\simeq |\psi\!>(E-M)^{-1}<\!\!\psi|+regular\; terms$, while $|\psi>$ are state vectors
carrying the quantum numbers of the resonance and $M=m_\ast-i\,\Gamma /2$ is the
complex mass of the resonance ($m_\ast$ is the real part of the mass $M$, $\Gamma$ is the
resonance width). 

To preserve unitarity, one has to introduce two distinct field operators
$\bar{\Psi}_L (x)$ and $\Psi_R (x)$ (``left'' and ``right'' eigen-fields) and
their complex conjugates, to describe one resonance degree of freedom in an
effective field theoretic way by the following free Lagrangian density:
\begin{eqnarray} {\cal L}_0 (x) & = &
\alpha \; \bar{\Psi}_L (x) \;
\bigg( \frac{1}{2} \; (i \! \slsh{\partial} 
                  - i \!\! \stackrel{\leftarrow}{\slsh{\partial}} ) 
                  - m_\ast + \frac{i}{2} \, \Gamma
\bigg) \; \Psi_R (x)  \nonumber \\
 & & \nonumber \\
 & & + \; 
\alpha^\ast \; \bar{\Psi}_R (x) \;
\bigg( \frac{1}{2} \; (\underbrace{i \! \slsh{\partial} 
                  - i \!\! \stackrel{\leftarrow}{\slsh{\partial}}
                  }_{\displaystyle =: i \!\!
                  \stackrel{\;\,\leftrightarrow}{\slsh{\partial}}} ) 
                  - m_\ast - \frac{i}{2} \, \underbrace{\gamma_0 \, \Gamma^+
                  \, \gamma_0}_{\displaystyle \stackrel{!}{=} \Gamma}
\bigg) \; \Psi_L (x) \nonumber   
\end{eqnarray}
($\alpha$ is an arbitrary complex constant chosen to be 1). Variation of the
action with respect to $\Psi^+_L (x)$,
$\Psi^+_R (x)$, $\Psi_L (x)$ and $\Psi_R (x)$ yields the following
``Dirac-equations'':
\[
\begin{array}{rcr}
 {(i \! \slsh{\partial} - M ) \; \Psi_R (x) \quad = \quad 0}\, ,
 & \quad &
 {\bar{\Psi}_R (x) (- i \!\! \stackrel{\leftarrow}{\slsh{\partial}} - M^\ast )
 \quad = \quad 0}\, , \\
 {(i \! \slsh{\partial} - M^\ast ) \; \Psi_L (x) \quad = \quad 0} \, , 
 & \quad  &
 {\bar{\Psi}_L (x) (- i \!\! \stackrel{\leftarrow}{\slsh{\partial}} - M ) 
 \quad = \quad 0} \, .
\end{array}
\]
Quantization by introduction of the four canonical conjugate momenta results in
the following non-vanishing equal-time anti-commutation relations:
\[
{\{ \, \Psi_{R,\, \sigma} (\vec{x},t) , \Psi^+_{L,\, \tau} (\vec{y},t) \,
\} 
 \quad = \quad \delta^{\, 3} (\vec{x} - \vec{y}) \; \delta_{\sigma \tau}} 
\, , \quad \mbox{\& Hermitian conjugate}\, .
\]
With the help of
$\omega_R \, ( \oabs{k} ) := \sqrt{\oabsq{k} + M^2} $ and 
$ \omega_L \, ( \oabs{k} ) := \sqrt{\oabsq{k} + M^{\ast \, 2}} 
\stackrel{!}{=} \omega^\ast_R \, ( \oabs{k} ) $ 
the following four-momenta can be defined: $k^{\,\mu}_R :=
(\omega_R \, ( \oabs{k} ) , \, \vec{k} \,\, )$ and $k^{\,\mu}_L := (\omega_L \,
( \oabs{k} ) , \, \vec{k} \,\, )$.\ Now generalized ``Dirac-spinors'' can be
introduced fulfilling the following momentum space ``Dirac-equations'':
\begin{eqnarray} 
 (- \, \slsh{k}_R + M ) \; u_R \, (\vec{k} , s) \quad = \quad 0 \, , & 
 & 
 \bar{u}_R \, (\vec{k} , s) \; (- \, \slsh{k}_L + M^\ast ) \quad = \quad 0 \, ,
 \nonumber \\
 (- \, \slsh{k}_L + M^\ast ) \; u_L \, (\vec{k} , s) \quad = \quad 0 \, , &  & 
 \;\, \bar{u}_L \, (\vec{k} , s) \; (- \, \slsh{k}_R + M ) \quad = \quad 0 \, , \nonumber
 \\
 (\slsh{k}_R + M ) \; v_R \, (\vec{k} , s) \quad = \quad 0 \, , &  & 
 \;\;\;\,\bar{v}_R \, (\vec{k} , s) \; (\slsh{k}_L + M^\ast ) \quad = \quad 0 \, , \nonumber
 \\
 (\slsh{k}_L + M^\ast ) \; v_L \, (\vec{k} , s) \quad = \quad 0 \, , & 
 & 
  \;\;\;\;\;\bar{v}_L \, (\vec{k} , s) \; (\slsh{k}_R + M ) \quad = \quad 0 \, . \nonumber  
\end{eqnarray}
Some of their important properties are:
\begin{eqnarray} 
\bar{u}_L \, (\vec{k} , s) \; u_R \, (\vec{k} , s^\prime ) \quad = \quad
\delta_{ss^\prime} \, , \, 
 & \quad & 
\bar{v}_L \, (\vec{k} , s) \; v_R \, (\vec{k} , s^\prime ) \quad = \quad - \,
\delta_{ss^\prime} \, , \nonumber \\ 
\bar{u}_L \, (\vec{k} , s) \; v_R \, (\vec{k} , s^\prime ) \quad = \quad 0 \, , \quad
 & \quad & 
\bar{v}_L \, (\vec{k} , s) \; u_R \, (\vec{k} , s^\prime ) \quad = \quad 0 \, ,
\nonumber
\end{eqnarray}
\begin{eqnarray} 
u^+_L \, (\vec{k} , s) \; u_R \, (\vec{k} , s^\prime )  = 
 \frac{\omega_R \, ( \oabs{k} )}{M} \; \delta_{ss^\prime} \, ,
 & \quad  & 
v^+_L \, (\vec{k} , s) \; v_R \, (\vec{k} , s^\prime )  = 
 \frac{\omega_R \, ( \oabs{k} )}{M} \; \delta_{ss^\prime} \, , \nonumber 
\end{eqnarray}
\[ 
\sum\limits_{s} u_R \, (\vec{k} , s ) \; \bar{u}_L \, (\vec{k} , s) =  
\frac{\slsh{k}_R + M }{2 M} 
\, ,\quad  - \, \sum\limits_{s} v_R \, (\vec{k} , s ) \; \bar{v}_L \, (\vec{k} , s) 
=  
\frac{- \, \slsh{k}_R + M }{2 M} \, .
\]
Using$\,$these spinors the$\,\,$following$\,$Fourier-decomposition$\,\,$of$\,\,$the$\,$field 
operators is performed:
\begin{eqnarray}
\Psi_R (x) & = & \sum\limits_{s} \, \int \! \!
\frac{d^3 k \; \sqrt{2 M}}{\sqrt{(2\pi )^3 \; 2 \omega_R \, (\oabs{k} )}}
\Big[ 
 u_R \, (\vec{k} , s ) \; b_R \, (\vec{k} , s ) \; e^{\displaystyle - \, i k_R
 x} \nonumber \\
 & & \quad\qquad\qquad\qquad\qquad\qquad +
 v_R \, (\vec{k} , s ) \; d^+_R \, (\vec{k} , s ) \; e^{\displaystyle i k_R x}
\Big] \, , \nonumber \\
 & & \nonumber \\
\Psi_L (x) & = & \sum\limits_{s} \, \int \! \!
\frac{d^3 k \; \sqrt{2 M^\ast}}{\sqrt{(2\pi )^3 \; 2 \omega_L \, (\oabs{k} )}}
\Big[ 
 u_L \, (\vec{k} , s ) \; b_L \, (\vec{k} , s ) \; e^{\displaystyle - \, i k_L
 x} \nonumber \\
 & & \quad\qquad\qquad\qquad\qquad\qquad +
 v_L \, (\vec{k} , s ) \; d^+_L \, (\vec{k} , s ) \; e^{\displaystyle i k_L x}
\Big] \, , \nonumber \\
 & & \nonumber \\
\bar{\Psi}_R (x) & = & \sum\limits_{s} \, \int \! \!
\frac{d^3 k \; \sqrt{2 M^\ast}}{\sqrt{(2\pi )^3 \; 2 \omega_L \, (\oabs{k} )}}
\Big[ 
 \bar{v}_R \, (\vec{k} , s ) \; d_R \, (\vec{k} , s ) \; e^{\displaystyle - \,
 i k_L x} \nonumber \\
 & & \quad\qquad\qquad\qquad\qquad\qquad +
 \bar{u}_R \, (\vec{k} , s ) \; b^+_R \, (\vec{k} , s ) \; e^{\displaystyle i
 k_L x}
\Big] \, , \nonumber \\
 & & \nonumber \\
\bar{\Psi}_L (x) & = & \sum\limits_{s} \, \int \! \!
\frac{d^3 k \; \sqrt{2 M}}{\sqrt{(2\pi )^3 \; 2 \omega_R \, (\oabs{k} )}}
\Big[ 
 \bar{v}_L \, (\vec{k} , s ) \; d_L \, (\vec{k} , s ) \; e^{\displaystyle - \,
 i k_R x} \nonumber \\
 & & \quad\qquad\qquad\qquad\qquad\qquad +
 \bar{u}_L \, (\vec{k} , s ) \; b^+_L \, (\vec{k} , s ) \; e^{\displaystyle i
 k_R x}
\Big] \, . \nonumber 
\end{eqnarray}
The non-vanishing anti--commutators of the momentum-space creation and anihilation
operators are:
\begin{eqnarray}
\{ \, b_R \, (\vec{k} , s ) ,  \, b^+_L \, (\vec{k}^\prime , s^\prime )
\} 
 & = & \delta^{\, 3} (\vec{k} - \vec{k}^\prime ) \;
 \delta_{ss^\prime} \, , \nonumber \\ 
\{ \, d_L \, (\vec{k} , s ) ,  \, d^+_R \, (\vec{k}^\prime , s^\prime )
\} 
 & = & \delta^{\, 3} (\vec{k} - \vec{k}^\prime ) \;
 \delta_{ss^\prime} \, , \nonumber \\
 & & \mbox{\& Hermitian conjugate}\, . \nonumber 
\end{eqnarray}
It is now straight forward to construct the ``Feynman-propagators'' by:

\begin{eqnarray}
\lefteqn{i \, \Delta^{R}_{F} (x-y) \quad := \quad 
  <0|T\,({\Psi}_R (x) \bar{\Psi}_L (y))|0> } \nonumber \\
 & \stackrel{!}{=} & 
i \int \! \frac{d^4p}{(2\pi )^4} \, e^{\displaystyle -ip (x-y)}
\frac{1}{\slsh{p} - M} =
i \int \! \frac{d^4p}{(2\pi )^4} \, e^{\displaystyle -ip (x-y)}
\frac{1}{p^2 - M^2} \; (\slsh{p} + M) \, , \nonumber \\ 
 & & \nonumber \\
\lefteqn{i \, \Delta^{L}_{F} (y-x) \quad := \quad 
 - \; \gamma_0 \; {\left( <0|T\,({\Psi}_R (x) \bar{\Psi}_L (y))|0> \right) }^+
 \gamma_0} \nonumber \\
 & \stackrel{!}{=} & 
i \int \! \frac{d^4p}{(2\pi )^4} \, e^{\displaystyle -ip (y-x)}
\frac{1}{\slsh{p} - M^\ast} =  
i \int \! \frac{d^4p}{(2\pi )^4} \, e^{\displaystyle -ip (y-x)}
\frac{1}{p^2 - M^{\ast\, 2}} \; (\slsh{p} + M^\ast) \, , \nonumber  
\end{eqnarray}
fulfilling the following equations:
\begin{eqnarray} (i \! \slsh{\partial}_x - M ) \; \Delta^{R}_{F} (x-y) \; = \; \delta^{\,
4}
(x-y) & , &
 \Delta^{R}_{F} (y-x) (- i \!\! \stackrel{\leftarrow}{\slsh{\partial}}_x - M )
 \;\; = \;
 \delta^{\, 4} (y-x) \, , 
       \nonumber \\
(i \! \slsh{\partial}_x - M^\ast ) \; \Delta^{L}_{F} (x-y) \; = \;
      \delta^{\, 4} (x-y) 
 & , & 
 \Delta^{L}_{F} (y-x) (- i \!\! \stackrel{\leftarrow}{\slsh{\partial}}_x -
 M^\ast ) \; = \;
 \delta^{\, 4} (y-x) \, . \nonumber 
\end{eqnarray}

\section{Effective model for one nucleon, one resonance and one
meson}
The model under consideration can now be extended by introduction of new degrees
of freedom, e.g. the nucleon field $N(x)$ (proton, neutron) and one meson 
$\phi_i (x)$ (internal index $i$), to obtain the following Lagrangian:
\[ {\cal L} (x) \quad = \quad {\cal L}^0_{N,N_\ast} (x) \quad
+ \quad {\cal L}^0_\Phi (x) \quad +
\quad {\cal L}^{\mbox{int}} (x) \, , \]
\[ \quad\qquad {\cal L}^0_{N,N_\ast} (x) = \left( \bar{N} (x), \bar{N}^R_\ast (x), \bar{N}^L_\ast
(x)\right) \; {\cal M} (N,N_\ast) \; 
\left( \begin{array}{l} N (x) \\ 
                       N^R_\ast (x) \\ 
                       N^L_\ast (x) \end{array} \right) \, , \]
\begin{eqnarray} {\cal L}^{\mbox{int}} (x)
 & = & - \;
\left( N^+ (x), N^{R\,+}_\ast (x), N^{L\,+}_\ast (x) \right) \nonumber \\
 & &
\Bigg[ \Gamma^{\,i} (N,N_\ast) \, \Phi_i (x) + 
{\left( \Gamma^{\,i} (N,N_\ast) \right)}^+ \,\Phi^+_i (x)
\Bigg] \!\!
\left(\begin{array}{l} N (x) \\ 
                       N^R_\ast (x) \\ 
                       N^L_\ast (x)
\end{array}\right) \, , \nonumber 
\end{eqnarray}
with the following $3\times 3$ matrices of Dirac-structures/operators:
\begin{eqnarray} {\cal M} (N,N_\ast) & := & \left(\begin{array}{ccc} {
\left( \frac{i}{2} \!\!\stackrel{\;\leftrightarrow}{\slsh{\partial}} - m
\right) } & 0 & 0 \\
 0 & 0 & {\alpha^\ast 
\left( \frac{i}{2} \!\!\stackrel{\;\leftrightarrow}{\slsh{\partial}} - M^\ast
\right) } \\
0 & {\alpha \left( \frac{i}{2} \!\!\stackrel{\;\leftrightarrow}{\slsh{\partial}} - M
\right) } & 0
\end{array} \right) \, , \nonumber \\
 & & \nonumber \\
\Gamma^{\,i} (N,N_\ast) & := &
\left(\begin{array}{lcc} \frac{1}{2} \Gamma^{\,i}_{\Phi N\rightarrow N} & 0 & 0
\\
 \Gamma^{\,i}_{\Phi N\rightarrow N^R_\ast} & \frac{1}{2}\Gamma^{\,i}_{\Phi
 N^R_\ast\rightarrow N^R_\ast} & 0 \\
 \Gamma^{\,i}_{\Phi N\rightarrow N^L_\ast} & \Gamma^{\,i}_{\Phi
 N^R_\ast\rightarrow N^L_\ast} & 
 \frac{1}{2} \Gamma^{\,i}_{\Phi N^L_\ast\rightarrow N^L_\ast} 
\end{array}\right) \, , \nonumber 
\end{eqnarray}
$\Gamma^{\,i} (N,N_\ast)$ should be called ``vertex matrix'' containing all vertex
structures between the fields considered. Summation over the internal indices $i$
of the meson field is required. The transition to the non-unitary
Wigner-Weisskopf approximation is performed by setting $N^{R\,+}_\ast
(x)=N^{L}_\ast (x)=0$.
\section{Implication to coupling constants}
As an example the non local interaction Lagrangian between the nucleon, the pion and the
Roper-resonance looks as follows:
\begin{eqnarray}
{\cal L}_{\pi NP_{11}} (x) 
& = & \frac{f_{\pi NP^L_{11}}}{m_{\pi}} \;
\left( \bar{N}^L_{P_{11}} (x) \; \gamma_\mu \gamma_5 \; \vec{\tau} \; N (x) 
\right) \; \cdot \; \partial^\mu {\vec{\Phi}}_{\pi} (x) 
 \nonumber \\
& & + \;  
\frac{f_{\pi NP^R_{11}}}{m_{\pi}} \;
\left( \bar{N}^R_{P_{11}} (x) \; \gamma_\mu \gamma_5 \; \vec{\tau} \; N (x) 
\right) \; \cdot \; \partial^\mu {\vec{\Phi}}_{\pi} (x) 
\quad + \quad
\mbox{h.c.} \nonumber 
\end{eqnarray}
Assuming the pseudoscalar couplings $g_{\pi NP^L_{11}}$ and $g_{\pi NP^R_{11}}$
to be equal (arbitrary complex numbers), consistency within the model requires
the following relations between the pseudovector couplings:
\begin{eqnarray}
 & & \frac{f_{\pi NP^L_{11}}}{m_\pi} \quad = \quad  
\frac{g_{\pi NP^L_{11}}}{M_{P_{11}}+m_N}
 \, , \quad  
 \frac{f_{\pi NP^R_{11}}}{m_\pi} \quad = \quad  
\frac{g_{\pi NP^R_{11}}}{M^\ast_{P_{11}}+m_N} \, , \nonumber \\
 & & \frac{f_{\pi NP^R_{11}}}{m_\pi} \quad = \quad  
  \frac{f_{\pi NP^L_{11}}}{m_\pi} \; \frac{M_{P_{11}}+m_N}{M^\ast_{P_{11}}+m_N}    
 \, , \quad  
g_{\pi NP^R_{11}} \quad = \quad g_{\pi NP^L_{11}} \, .  \nonumber
\end{eqnarray}
Similar expressions hold for negative parity resonances, e.g. the 
$S_{11}(1535)$ resonance:
\begin{eqnarray}
 & & \frac{f_{\pi NS^L_{11}}}{m_\pi} \quad = \quad  
\frac{g_{\pi NS^L_{11}}}{M_{S_{11}}-m_N}
 \, , \qquad  
 \frac{f_{\pi NS^R_{11}}}{m_\pi} \quad = \quad  
\frac{g_{\pi NS^R_{11}}}{M^\ast_{S_{11}}-m_N} \, , \nonumber \\
 & & \nonumber \\
 & & \frac{f_{\pi NS^R_{11}}}{m_\pi} \quad = \quad  
  \frac{f_{\pi NS^L_{11}}}{m_\pi} \; \frac{M_{S_{11}}-m_N}{M^\ast_{S_{11}}-m_N}   
 \, , \qquad  
g_{\pi NS^R_{11}} \quad = \quad g_{\pi NS^L_{11}} \, . \nonumber 
\end{eqnarray}

Obviously the ``left'' and
``right'' pseudovector couplings differ by complex phases which are determined by
the resonance width. {\em These phases are relevant to interference terms, which are
important for calculating threshold meson production processes at high momentum
transfers, in which different resonances are excited by one collision of e.g.\ protons
and nuclei ($\rightarrow\;$corre-lations)}. One process, which is very sensitive
to interference terms, is the proton induced $\eta$-production at threshold. As
this process is dominated by the excitation of just one resonance
($S_{11}(1535)$), the phases discussed in the model above can 
only be measured, if the resonance is excited by a nonlocal interaction and deexcited by a local interaction (and {\em vice versa}), or if the selfenergy of the resonance is treated to be not constant, but energy-dependent.  
A remarkable feature of the full effective field theoretic model is that
unitarity is guaranteed, although the coupling constants can have arbitrary 
{\em complex} values, which are related to the self-energies of the resonances.
This property should be not too surprising, as from renormalization theory it is
well known, that not only the mass has to be renormalized, but also the coupling
constants.
A special feature of resonance effective degrees of freedom is, that not only
the self energies are complex, but also the couplings. It is not clear, whether
a bosonization procedure applied to an elementary non abelian theory like QCD
will lead only to effective degrees of freedom, which have real self energies
like the nucleon. {\em The presented model shows, that resonant degrees of freedom
are also compatible with the requirement of unitarity}. 
\newpage
Finally one should mention a very subtle point for discussion, which has to be
solved consistently within such an effective, non local field theoretical model:
as the interaction between nucleons and mesons can generate resonances as poles of
the S-matrix in the complex energy plane, one has to make clear -- {\em to avoid double
counting} --, in what way the effective resonant degrees of freedom in the model above
have to be interpreted.

A detailed discussion of the full model will be given in [2]. The first, rough introduction can be found in [3].

\end{document}